\documentclass[10pt,a4paper,twoside]{article}
\usepackage{epsfig}
\usepackage{baltlat5}
\usepackage{wrapfig}
\begin{document}
\ \
\vspace{-0.5mm}

\setcounter{page}{301}
\vspace{-2mm}

\titlehead{Baltic Astronomy, vol.\ts 17, 301 (2008).}

\titleb{StH$\alpha$-55: A CARBON MIRA, NOT A SYMBIOTIC BINARY}

\begin{authorl}
\authorb{U.~Munari}{1}
\authorb{A.~Siviero}{1}
\authorb{M.~Graziani}{2}
\authorb{A.~Maitan}{2}
\authorb{A.~Henden}{3}
\authorb{L.~Baldinelli}{2}
\authorb{S.~Moretti}{2}
\authorb{S.~Tomaselli}{2}
\end{authorl}

\begin{addressl}

\addressb{1}{INAF Osservatorio Astronomico di Padova, via dell'Osservatorio
8, 36012 Asiago (VI), Italy}

\addressb{2}{ANS Collaboration, c/o Osservatorio Astronomico, via
dell'Osservatorio 8, 36012 Asiago (VI), Italy}

\addressb{3}{AAVSO, 49 Bay State Road, Cambridge, MA, USA}

\end{addressl}

\submitb{Received 2008 November 1; accepted 2008 December 20}

\begin{summary}

We carried out a $V$$R_{\rm C}$$I_{\rm C}$ photometric monitoring of
StH$\alpha$~55, and in addition we obtained low resolution absolute
spectro-photometry and high resolution Echelle spectroscopy. Our data show
that StH$\alpha$~55 is a carbon Mira, pulsating with a 395~day period, with
$<$$V$$>$=13.1 mean brightness and $\Delta V$=2.8~mag amplitude. It suffers
from a low reddening ($E_{B-V}$=0.15), lies at a distance of 5~kpc from the
Sun and 1~kpc from the galactic plane, and its heliocentric systemic
velocity is close to +22~km~sec$^{-1}$. The difference between the radial
velocity of the optical absorption spectrum and that of the H$\alpha$
emission is unusually small for a carbon Mira. The spectrum of
StH$\alpha$~55 can be classified as C-N5 C$_2$6$^{-}$. Its $^{13}$C/$^{12}$C
isotopic ratio is normal, and lines of BaII and other $s$-type elements, as
well as LiI, have the same intensity as in field carbon stars of similar
spectral type. The Balmer emission lines are very sharp and unlike those
seen in symbiotic binaries. Their intensity changes in phase with the
pulsation cycles in the same way as seen in field carbon Miras. We therefore
conclude that StH$\alpha$~55 is a bona fide, normal carbon Mira showing no
feature supporting a symbiotic binary classification, as previously
hypothesized.

\end{summary}


\sectionb{1}{INTRODUCTION}

Very scanty information is available in literature on StH$\alpha$~55. It was
discovered by Stephenson (1986) as a V$\sim$13.5~mag stellar source
displaying H$\alpha$ in emission. A low-resolution spectrum of
StH$\alpha$~55 was obtained by Downes and Keyes (1988), who classified it as
a carbon star and confirmed the presence of H$\alpha$ and H$\beta$ in
emission.  The star was then long forgotten until Belczynski et al. (2000)
inserted it in their Catalog of Symbiotic Stars. They cataloged it as a
suspected symbiotic star, arguing that the intensity of the H$\alpha$
emission on the spectrum presented by Downes and Keyes (1988) was larger
than in normal, single carbon stars. A second spectrum of StH$\alpha$~55 was
presented by Munari and Zwitter (2002) as part of their multi-epoch
spectro-photometric atlas of symbiotic stars, which surveyed the vast
majority of the sources listed by Belczynski et al. (2000). The Munari and
Zwitter (2002) spectrum of StH$\alpha$~55 was virtually identical to that of
Downes and Keyes (1988), arguing in favor of very modest, if any,
spectroscopic variability.

The rarity of galactic symbiotic stars containing a carbon donor star
prompted us to insert StH$\alpha$~55 among the $\sim$80 symbiotic stars that
the ANS (Asiago Novae and Symbiotic star) Collaboration is monitoring
spectroscopically and photometrically (UBVRcIc bands). In this paper we 
report observations and analysis that show StH$\alpha$~55 to be a
normal carbon Mira with no hint of a symbiotic binary nature.

    \begin{table}
      \centering
      \caption{Magnitude and colors of the photometric comparison stars.}
      \includegraphics[width=10.5cm]{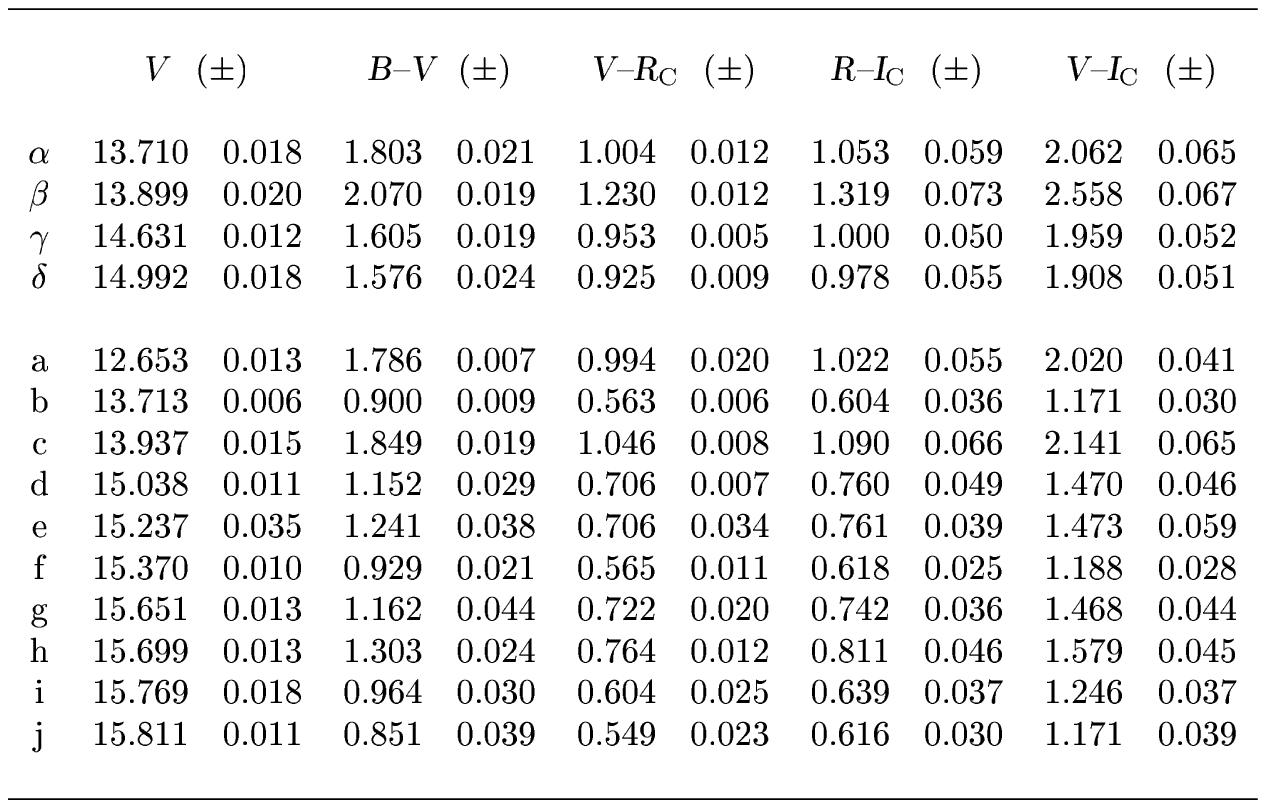}
    \end{table}

    \begin{table}
      \centering
      \caption{Magnitude and colors of StH$\alpha$~55 from our observations.}
      \includegraphics[width=12cm]{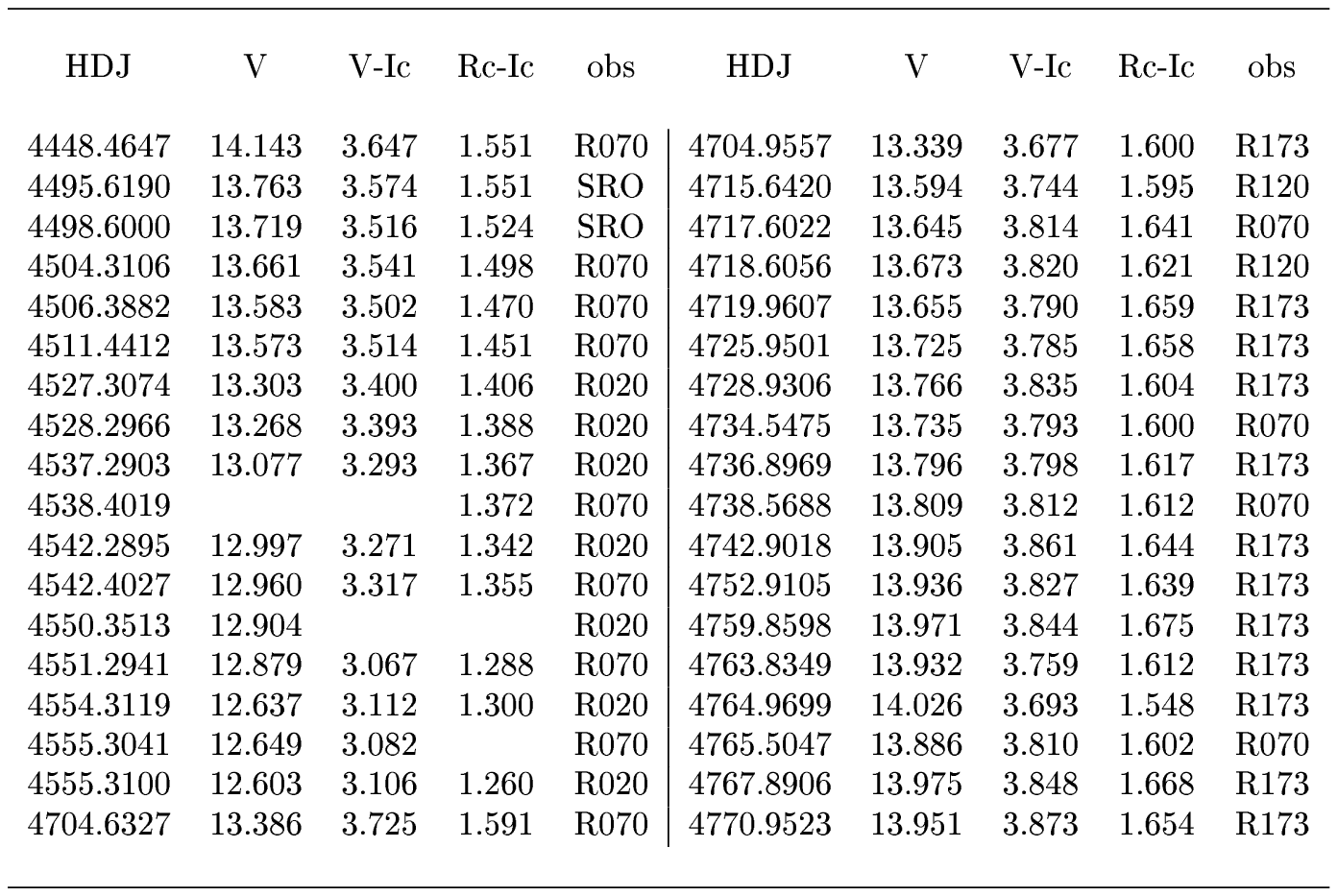}
    \end{table}

\sectionb{2}{OBSERVATIONS}
 
\subsectionb{2.1}{Photometry}

As a first step, a BVRcIc photometric sequence was calibrated around
StH$\alpha$~55 for use by all of the telescopes participating in the
monitoring effort. The sequence is presented in Table~1 and identified in
the finding chart of Figure~1. The sequence was calibrated against Landolt
(1983, 1992) equatorial standards with observations obtained with the 0.35m
robotic telescope of Sonoita Research Observatory (SRO, Arizona, USA). It uses
$B$$V$$R_{\rm C}$$I_{\rm C}$ Optec filters and an SBIG STL-1001E CCD camera,
1024$\times$1024 array, 24 $\mu$m pixels $\equiv$ 1.25$^{\prime\prime}$/pix,
with a field of view of 20$^\prime$$\times$20$^\prime$.

All our subsequent observations of StH$\alpha$~55 were accurately reduced 
and color-corrected against the sequence in Table~1. The results are listed in
Table~2. Typical global errors (including quadratically the contribution of
Poissonian noise, color transformations and residuals in the dark/flat/bias
corrections) are 0.035~mag in V, 0.027 in V-Ic, and 0.018 in Rc-Ic. The
observations of Table~2 were collected with the following ANS Collaboration
telescopes. {\em R020}:
the 0.40-m f/5 Newton reflector of the P.~Pizzinato Observatory located in
Bologna (Italy). The CCD is a HiSis~23ME, 768$\times$1157 array, 9~$\mu$m
pixels $\equiv$ 0.62$^{\prime\prime}$/pix, with a field of view of
8$^\prime$$\times$12$^\prime$. The $B$$V$$R_{\rm C}$$I_{\rm C}$ filters are
from Schuler. {\em R070}: a 0.30-m f/10 Meade LX200 telescope privately
owned by one of us (M.G.) and operated in Alfonsine (RA, Italy). It is equipped with
$B$$V$$R_{\rm C}$$I_{\rm C}$ Schuler filters. The CCD is a Finger Lake
Instruments ML0261E, 512$\times$512 array, 20 $\mu$m pixels $\equiv$
1.35$^{\prime\prime}$/pix, with a field of view of
11$^\prime$$\times$11$^\prime$. {\em R120}: the 0.42-m f/5.4 Newton
telescope operated in Bastia (Ravenna, Italy) by Associazione Ravennate
Astrofili Rheyta. It has an Apogee Alta 260e CCD Camera, 512$\times$512
array, 20 $\mu$m pixels $\equiv$ 1.83$^{\prime\prime}$/pix, for a field of
view of 16$^\prime$$\times$16$^\prime$. It is used in combination with
Schuler $U$$B$$V$$R_{\rm C}$$I_{\rm C}$ filters. {\em R173}: a 0.30-m f/11.9
Dall-Kirkham robotic telescope, part of the GRAS network (GRAS1, New Mexico,
USA). It carries a Finger Lake Instruments IMG1024 DM CCD camera,
1024$\times$1024 array, 24 $\mu$m pixels $\equiv$ 1.38$^{\prime\prime}$/pix,
with a field of view of 24$^\prime$$\times$24$^\prime$; The $B$$V$$R_{\rm
C}$$I_{\rm C}$ filters are from Optec.

\subsectionb{2.2}{Spectroscopy}

A low resolution, absolutely fluxed spectrum of  StH$\alpha$~55  was obtained on
February 25, 2008 with the B\&C spectrograph of the INAF Astronomical
Observatory of Padova attached to the 1.22m telescope operated in Asiago by
the Department of Astronomy of the University of Padova. The slit, aligned
with the parallactic angle, had a 2 arcsec sky projection, and the
total exposure time was 55~minutes. The detector was an ANDOR iDus 440A CCD
camera, equipped with a EEV 42-10BU back illuminated chip, 2048$\times$512
pixels of 13.5~$\mu$m size. A 300 ln/mm grating blazed at 5000~\AA\ provided
a dispersion of 2.26~\AA/pix and a wavelength range extending from 3400 to
8100~\AA.

A high resolution spectra of StH$\alpha$~55 was obtained on 20 March 2008
with the Echelle spectrograph mounted on the 1.82m telescope operated in
Asiago by INAF Astronomical Observatory of Padova. The detector was an
EEV~CCD47-10 CCD, 1024$\times$1024 array, 13 $\mu$m pixel, covering the
interval 3600$-$7300~\AA\ in 31 orders. A slit width of 200~$\mu$m provided
a resolving power $R_P$=22\,000.

\begin{figure}
\centerline{\psfig{figure=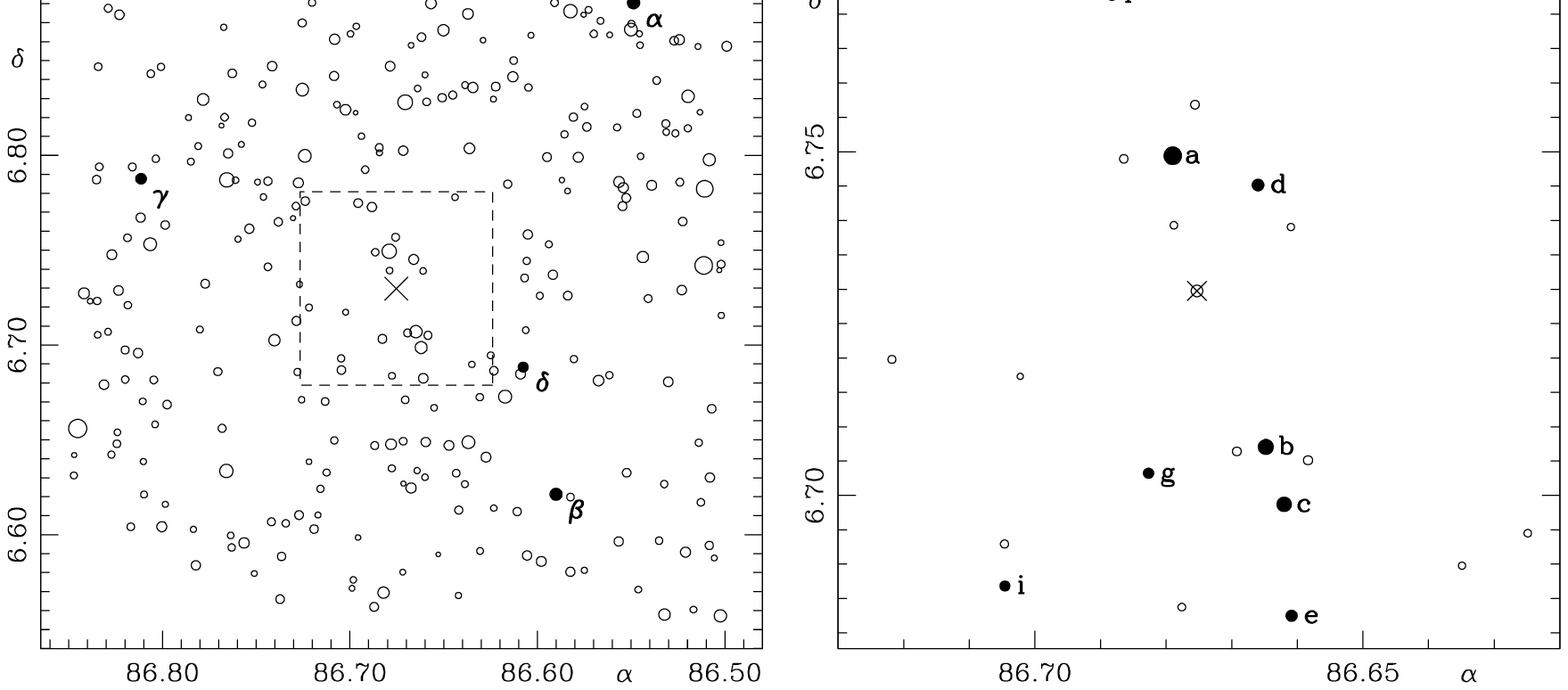,width=12.5truecm,angle=0,clip=}}
\captionc{1}{Identification charts for the photometric comparison stars
around StH$\alpha$~55 listed in Table~1.}
\end{figure}

\begin{figure}
\centerline{\psfig{figure=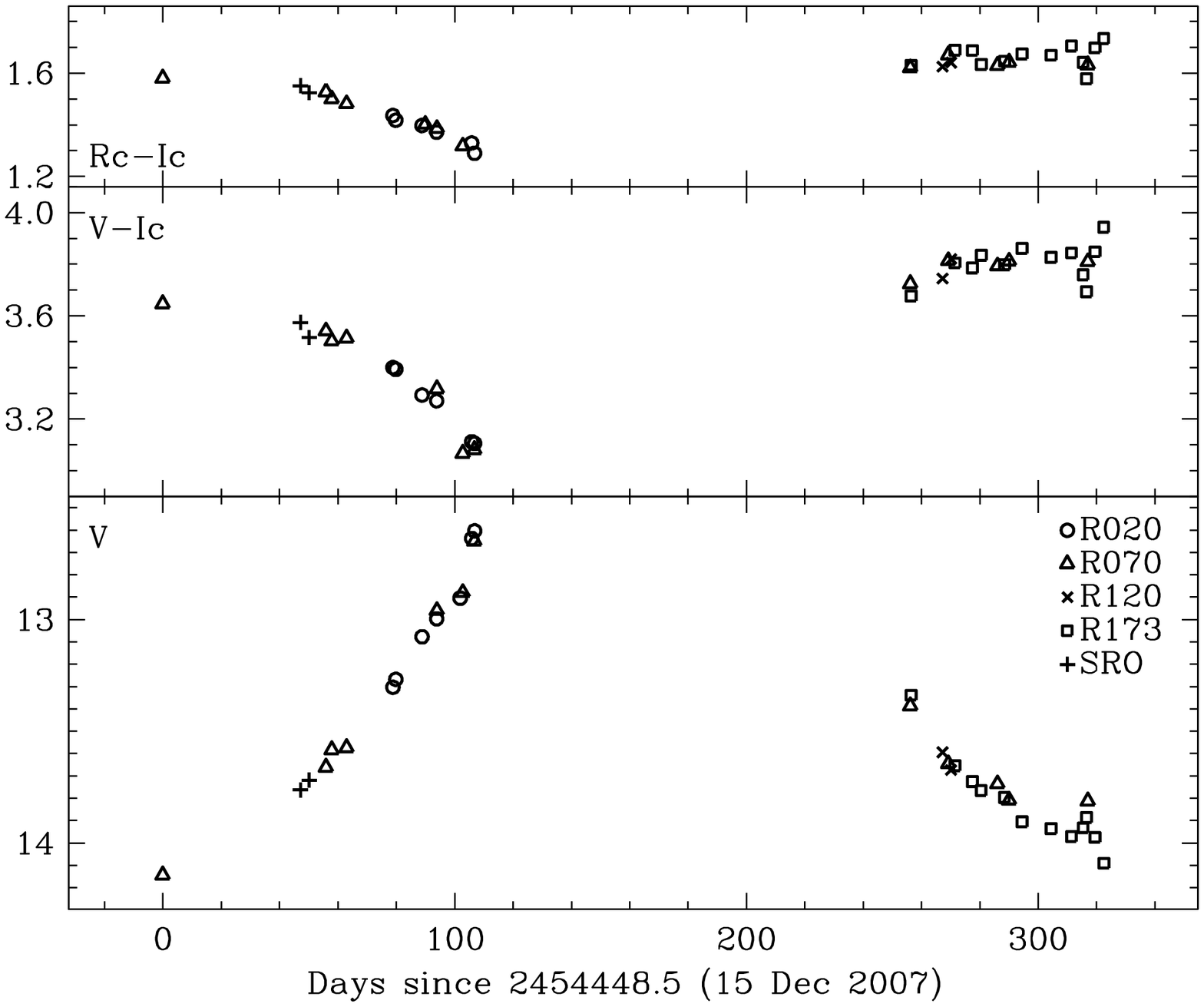,width=10.5truecm,angle=0,clip=}}
\captionc{2}{Light- and color-curves of StH$\alpha$~55 from the data in Table~1.
             For instrument identification, see Section 2.1.}
\end{figure}

\sectionb{3}{RESULTS}

\subsectionb{3.1}{A carbon Mira}

The light-curve presented in Figure~2 shows that StH$\alpha$~55 is
indeed variable, with a large amplitude, a long period, and bluer colors
when the star is brighter and redder colors when it is fainter. These are 
distinctive features of Mira-like pulsations.

To derive the pulsation period, we have searched external data archives
for additional observations that would fill in the gaps of our photometric
monitoring. We found additional V-band data in the ASAS database (All Sky
Automated Survey, Pojmanski 1997), covering the time interval from 18
January 2006 to 27 April 2006. We carried out a Deeming-Fourier
(Deeming 1975) period search on our V-band set combined with that from ASAS,
which resulted in a clear and strong 395~day periodicity. 
The combined V-band data are phase plotted in Figure~3 according to the
ephemeris:
\begin{equation}
Max(V) = 2453849 + 395 \times E
\end{equation}
The lightcurve in Figure~3 is that of a Mira variable, with $<$$V$$>$=13.1
mean brightness and $\Delta V$=2.8~mag amplitude, that varies between
V=14.28 and V=11.48~mag. The lightcurve appears quite symmetric, the
pulsation period is rather long and the amplitude is small, all distinctive
characteristics of carbon Miras in comparison with their O-rich counterparts
(Mennessier et al. 2000).

The lightcurve in Figure~2 shows a short-lasting departure from a smooth
trend around day 315. This part of the lightcurve is magnified in Figure~4,
which shows that the event lasted about a week and was
characterized by a $\Delta V$=0.1~mag brightening accompanied by a
simultaneous $\Delta(V-I_{\rm C})$$\sim$0.2~mag, $\Delta(R_{\rm C}-I_{\rm
C})$$\sim$0.15~mag blueing of the colors.
This occurrence briefly interrupted the smooth fading of the lightcurve
towards minimum. Its isolated occurrence makes its
interpretation rather speculative. For sake of discussion, it could be
argued that the event traced either a short-lived halt in the expansion of
the atmosphere of the Mira toward minimum brightness (corresponding to 
the largest radius), or the emergence of a convection cell hotter than the
surrounding stellar surface.

\begin{figure}
\centerline{\psfig{figure=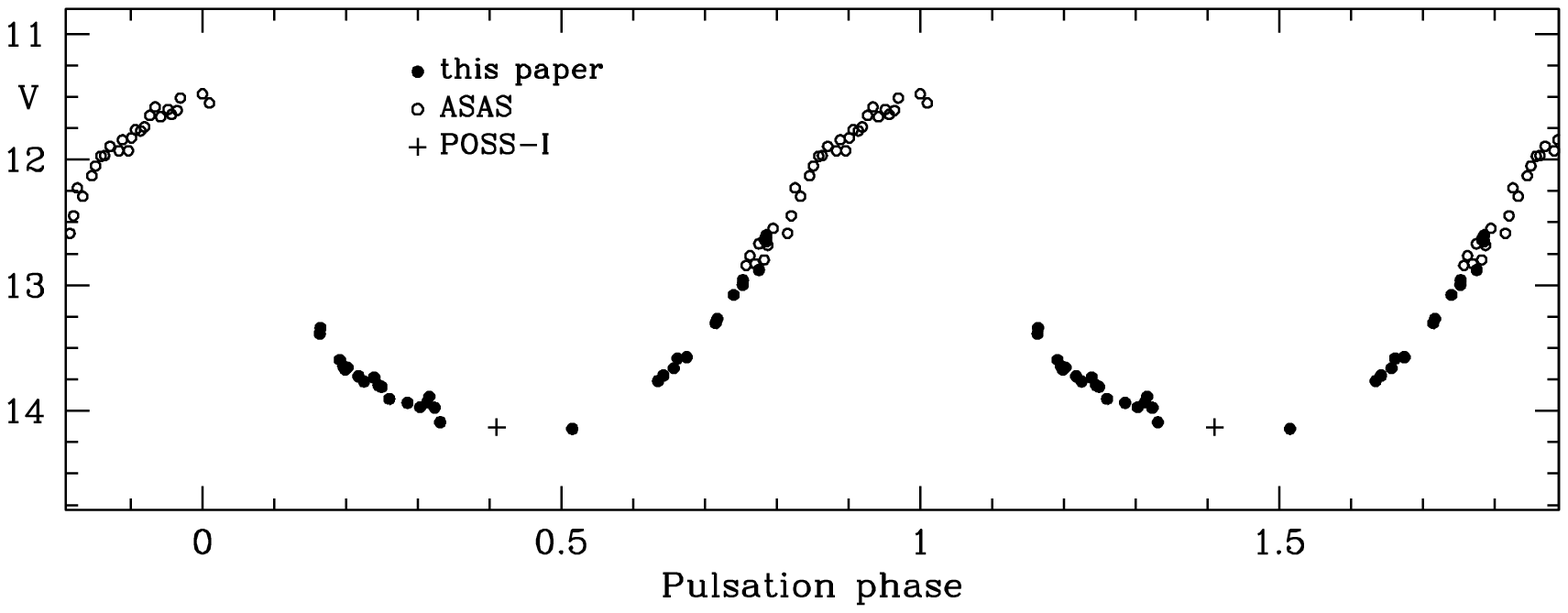,width=12truecm,angle=0,clip=}}
\captionc{3}{Plot of our (from Table~1, solid circles) and
             ASAS (open circles) V-band data, phased according 
             to ephemeris (1).}  
\end{figure}

\subsectionb{3.3}{Carbon star classification}

The spectrum of StH$\alpha$~55 presented in Figure~5 is that of a normal
carbon star. A comparison with the Barnbaum et al. (1996) spectral atlas
suggests a classification on the revised MK system of Keenan (1993) as
C-N5~C$_2$6$^{-}$, i.e. among the coolest and C-richest carbon stars.  The
$^{13}$C/$^{12}$C isotopic ratio is normal, as derived by the comparison of
the strength of the 6260~\AA\ band of $^{13}$C$^{14}$N with the 6206~\AA\
band of $^{12}$C$^{14}$N. Lines of BaII and other $s$-type elements, as well
as LiI, have the same intensity in StH$\alpha$~55 as in field carbon stars
of the same spectral type (see for ex. Jaschek and Jaschek 1987, 
Wallerstein and Knapp 1998).

\subsectionb{3.4}{Reddening}

Feast et al. (1990) proposed the following statistical expression for the
reddening as function of distance and galactic $b$ latitude:
$E_{B-V}=0.032({\rm cosec} |b| - 1)[1 - exp(-10r\sin |b|)]$, where $r$ is
the distance in kpc.  For the 5~kpc distance derived in next section, this
relation provides $E_{B-V}$=0.13 for StH$\alpha$~55. The three-dimensional
mapping of the galactic interstellar extinction by Arenou et al. (1992),
gives for the direction and distance to StH$\alpha$~55 the value $A_V$=0.53,
and thus a quite similar reddening $E_{B-V}$=0.17. Therefore, in this
paper we adopt $E_{B-V}$=0.15 as the interstellar reddening affecting
StH$\alpha$~55. 

\subsectionb{3.5}{Distance}

\begin{wrapfigure}[21]{l}[0pt]{65mm}
\vbox{
\centerline{\psfig{figure=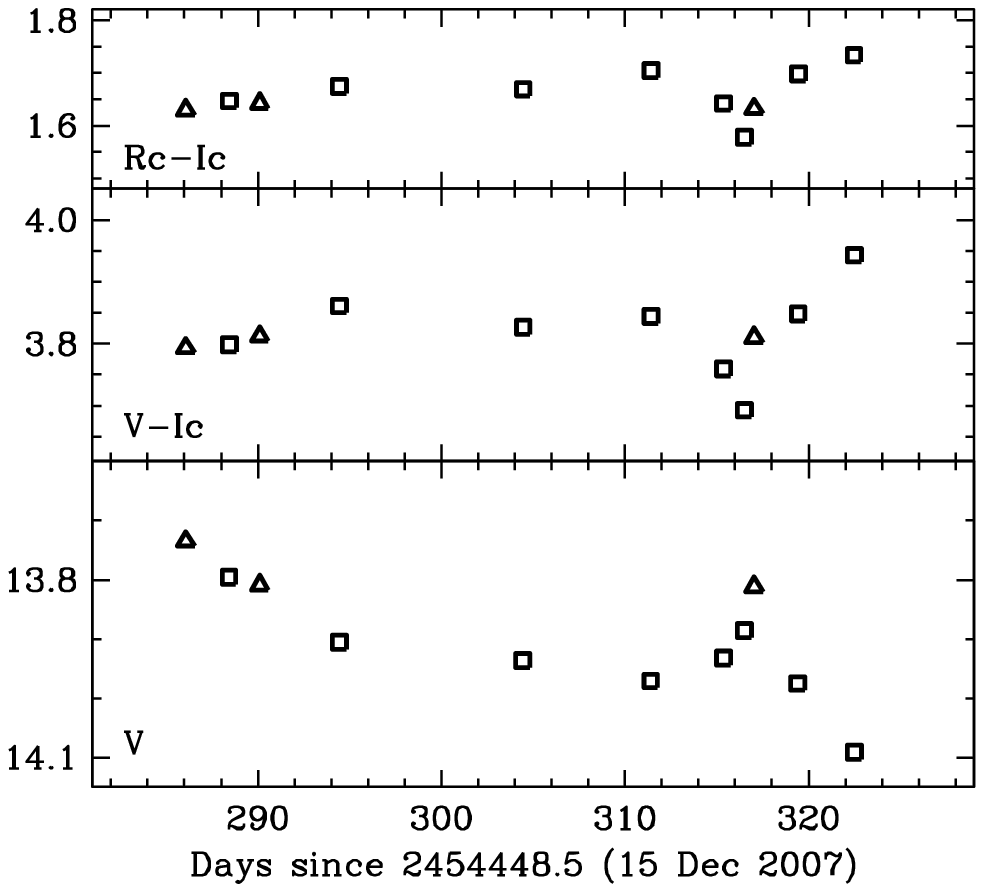,width=63truemm,angle=0,clip=}}
\captionc{4}{A portion of the lightcurve in Figure~2 that includes the
brightness glitch of day  $\sim$315.}
}
\end{wrapfigure}

The most recent calibration of the various period-luminosity relations
applicable to RGB and AGB variables can be found in Soszynski et al.
(2007). Their relation for C-rich Mira variables in LMC is $W_I = -6.618\log
^2 P + 25.468 \log P -12.522$, where $W_I=I_{\rm C} - 1.55(V-I_{\rm C})$ is
a reddening free index. The mean values for StH$\alpha$~55 are $<$$I_{\rm
C}$$>$=9.7 and $<$$V - I_{\rm C}$$>$=3.4, that correspond to an observed
$W_I^{obs}$=+4.43. The value computed for a 395~day pulsation period is
$W_I^{calc}$=+8.94. Adopting a LMC distance modulus of ($m$ -
M)$_\circ$=18.39 (van Leeuwen et al. 2007), a LMC reddening of
$E_{B-V}$=0.06 (Mateo 1998), the extinction relation $A_{\rm Ic} = 2.1
\times E_{B-V}$ (valid for a cool spectral distribution and a standard
$R_V$=3.1 extinction law, Fiorucci and Munari 2003), and scaling to solar
metallicity (following Soszynski et al. 2007), the distance to
StH$\alpha$~55 is 5.2~kpc.

The 2MASS survey measured StH$\alpha$~55 at $J$= 8.114$\pm$0.034, $H$=
6.507$\pm$0.034 and $K_s$= 5.297$\pm$0.023~mag on JD 2451458.8462, which
corresponds to a pulsation phase 0.98 according to Eq.(1). This translates
into a pulsation phase 0.88 in the infrared, where the maximum occurs about
0.1 phases later than in the optical. O-rich Miras of 395~days period have
amplitudes in the $K$ band of the order of 0.85~mag (Whitelock et al. 1991),
and their C-rich counterparts have distinctively lower amplitudes
$\Delta K$$\approx$0.6~mag (Whitelock et al. 2006).
Because the $K$-band lightcurve of carbon Miras are closely sinusoidal in
shape (Kerschbaum et al. 2006), the 2MASS $K_s$= 5.297 at phase 0.88 would 
translate into a mean $K_s$ brightness of StH$\alpha$~55 of $<$$K_s$$>$=5.297 
+ 0.3$\times$0.73$\approx$5.52~mag. Whitelock et al. (2008) have calibrated,
on the revised Hipparcos parallaxes by van~Leeuwen (2007), the following
period-luminosity relation for Galactic carbon Miras: $M_K = -3.52(\pm
0.36)[\log P - 2.38] - 7.24 (\pm0.07)$. Ignoring the difference between the
$K_s$ and $K$ bands (cf. Tokunaga et al. 2002), the distance to
StH$\alpha$~55 would be 5.0~kpc for the $E_{B-V}$=0.15 above derived. This
is remarkably close to the distance obtained above from the reddening-free
Weisenheit $W_I$ index.

\begin{figure}
\centerline{\psfig{figure=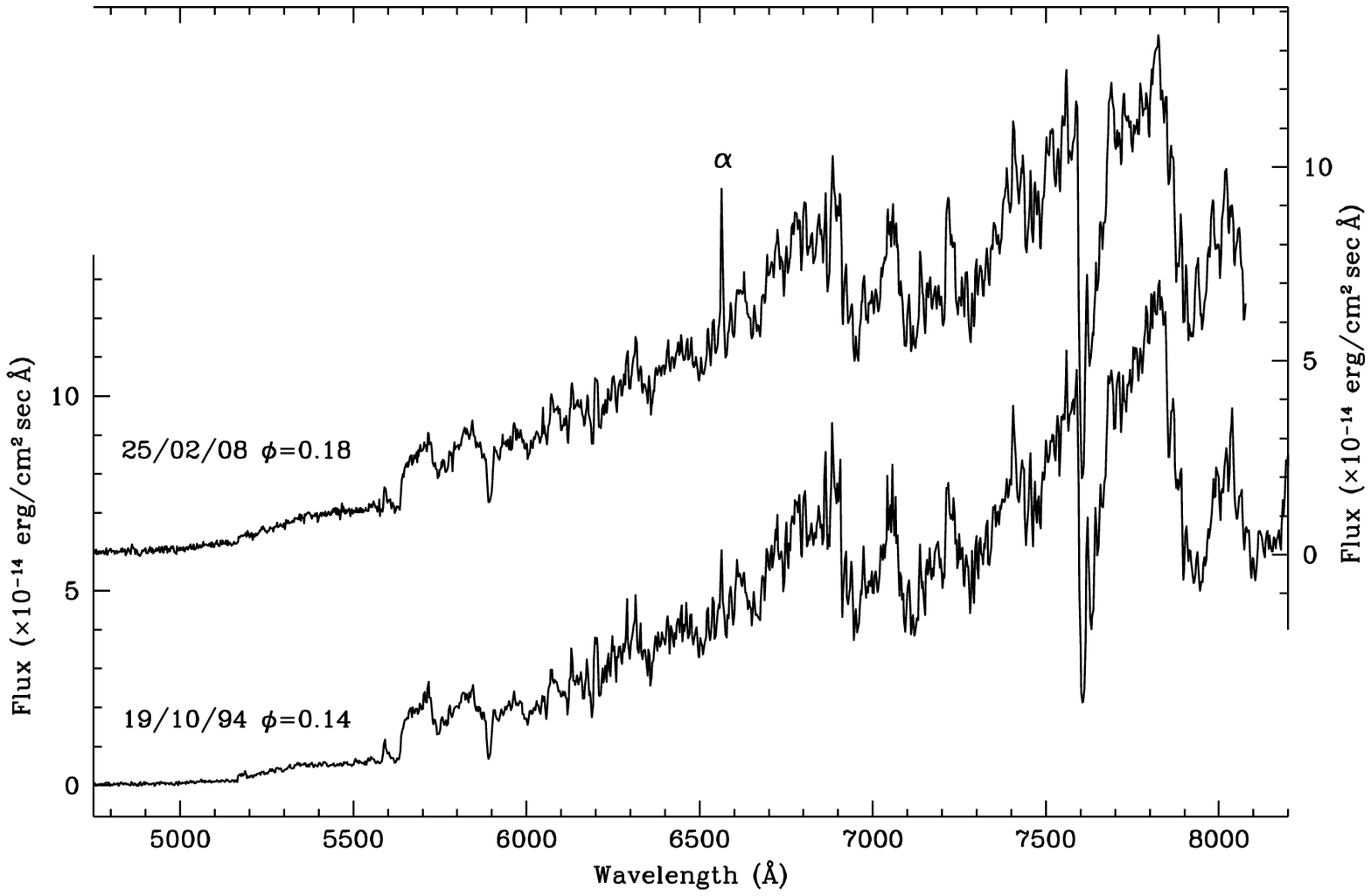,width=125truemm,angle=0,clip=}}
\captionc{5}{Comparison of Asiago 1.22m and ESO 1.5m low resolution, absolutely
             fluxed spectra of StH$\alpha$~55. $\phi$ indicates the pulsation phase
             according to ephemeris (1).}
\end{figure}

In this paper we therefore adopt a 5~kpc distance to StH$\alpha$~55. It
agrees with the lack of significant proper motion of StH$\alpha$~55, as listed
by the NOMAD database. At Galactic coordinates $l$=199.3 and $b$=$-$11.1,
the corresponding height above the Galactic plane of StH$\alpha$~55 is
$z$=1~kpc. It is quite far from the 190~pc scale height of N-type carbon
stars found by Dean (1976) and the 180~pc scale height of Galactic carbon
Miras found by Kerschbaum and Hron (1992). It suggests a possible
association of StH$\alpha$~55 with the Thick disk/inner Halo of the Galaxy
and not with the Thin disk with which N-type carbon stars are usually
associated (Keenan 1993). Using the results of Feast et al. (2006),
the 395~day period would correspond to a 2.7~Gyr age and initial    
1.6~M$_\odot$ mass for StH$\alpha$~55.

\subsectionb{3.5}{Radial velocity}

\begin{wrapfigure}[19]{l}[0pt]{70mm}
\vbox{
\centerline{\psfig{figure=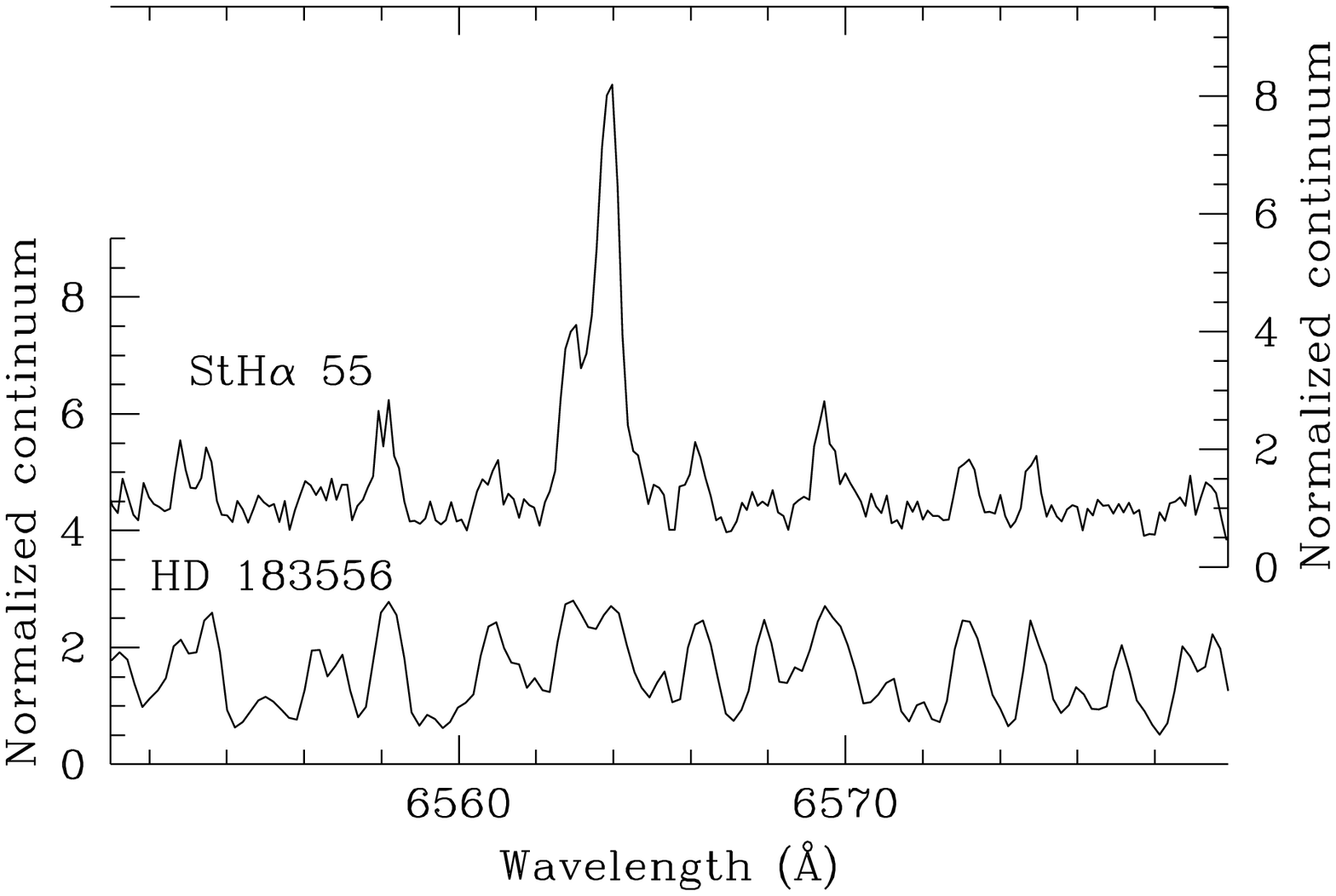,width=70truemm,angle=0,clip=}}
\captionc{6}{Comparison of the region around H$\alpha$ from Asiago
1.82m Echelle spectra of StH$\alpha$~55 and of the non-Mira, C~6-4 carbon
star HD~183556.}
}
\end{wrapfigure}

The heliocentric radial velocity of StH$\alpha$~55 on the Echelle spectrum is
+23.7 ($\pm$0.2) km~sec$^{-1}$, obtaine by cross-correlation with the appropriate
spectrum from the synthetic spectral library of carbon stars of Pavlenko et
al. (2003). The spectrum was obtained at pulsation phase 0.52, thus at
minimum brightness, when the radial velocity normally associated with the
Mira pulsation reaches its minimum velocity (cf Joy 1954, Hoffmeister et al.
1985). The typical radial velocity amplitude at optical wavelengths of
carbon Miras is $\sim$8~km~sec$^{-1}$ (Sanford 1950, Barnbaum 1992a), and
therefore we could conclude that the optical mean velocity of StH$\alpha$~55
should be near +28 km~sec$^{-1}$. It is known that the optical mean velocity
of Carbon Miras is offset by some km~sec$^{-1}$ from the barycentric
velocity (better traced by CO infrared observations, Nowotny et al. 2005). 
Barnbaum (1992b) and Barnbaum and Hinkle (1995) find that barycentric
velocities of Carbon Miras are on average blueshifted by
$\sim$6~km~sec$^{-1}$ from mean optical velocities. This would translate
into a heliocentric systemic velocity of +22 km~sec$^{-1}$ for
StH$\alpha$~55.

The heliocentric radial velocity of the H$\alpha$ emission line is
+19.3 ($\pm$0.6) km~sec$^{-1}$, i.e. blue-shifted by 4.4 km~sec$^{-1}$
with respect to the Mira absorption spectrum. It is known that the velocity
of the emission lines differs from that of the aborption spectrum in 
both oxygen- and carbon-rich Miras. The difference for carbon Miras is
reported as $<$RV$_{em}^{H\alpha}$ - RV$_{abs}$$>$ = $-$30~km~sec$^{-1}$
according to Menzies et al. (2006) and $-$20km~sec$^{-1}$ following
Sanford (1944), with minimal dispersions around the respective means. 
In none of the 43 stars analyzed by  Menzies et al.,
or in the 34 stars studied by Sanford, is the difference between
the radial velocity of the absorption spectrum and that of the
H$\alpha$ emission as small as it is in StH$\alpha$~55 .

\subsectionb{3.6}{Profile and Variability of H$\alpha$ emission}

The main reason for Belczynski et al. (2000) to include StH$\alpha$~55 in
their list of suspected symbiotic stars was the ''HI emission lines too
strong for a single carbon star''. Figure~6 shows a portion of our Echelle
spectrum of StH$\alpha$~55 centered on H$\alpha$, and for comparison the
equivalent portion of a spectrum of the non-variable carbon star HD~183556
obtained with the Asiago Echelle spectrograph and the same instrumental
set-up as adopted for StH$\alpha$~55.

\begin{wrapfigure}[27]{l}[0pt]{70mm}
\vbox{
\centerline{\psfig{figure=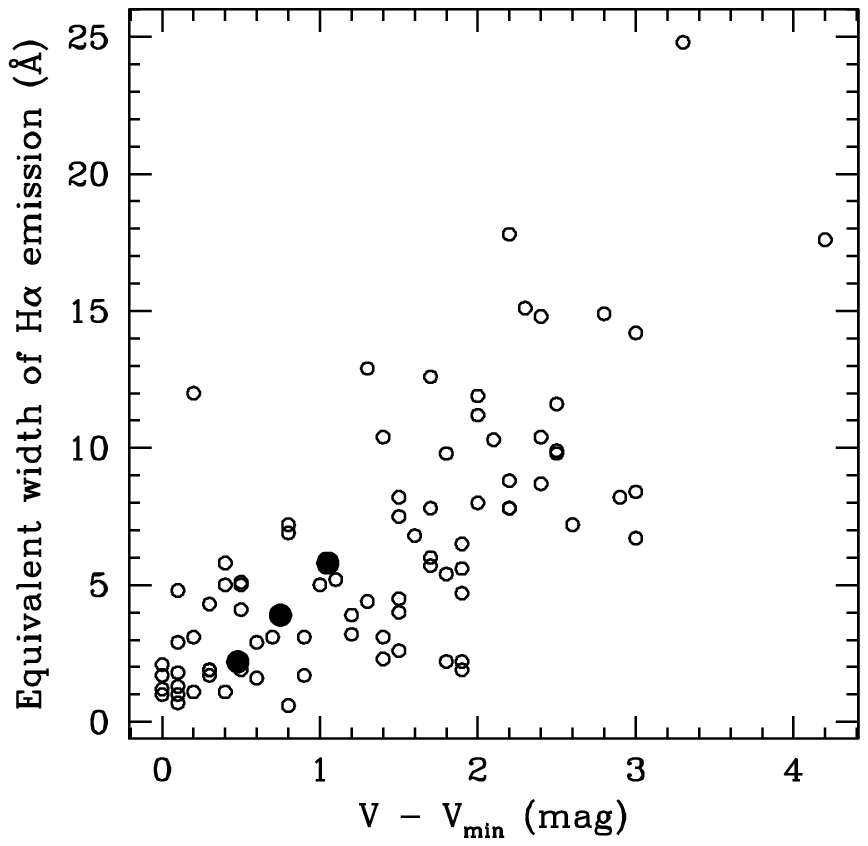,width=70truemm,angle=0,clip=}}
\captionc{7}{Relation between the equivalent width of H$\alpha$ (in \AA)
and magnitude of carbon Miras (from Mikulasek and Graf 2005, open circles).
The solid points represents our measurements of the equivalent width of H$\alpha$ 
emission in StH$\alpha$~55.}
}
\end{wrapfigure}

The H$\alpha$ emission component of StH$\alpha$~55 is sharp and very similar
to those typically observed in carbon Miras, as nicely seen in comparison
with the high resolution H$\alpha$ spectral atlas of Mikulasek and Graf
(2005). On the contrary, the H$\alpha$ emission lines of symbiotic stars are
far wider and present multi-components and frequent central reversals (e.g.
Ivison et al. 1994, van Winckel et al. 1993). Carbon symbiotic stars are no
exception, as illustrated by the Asiago Echelle observations of Munari
(1991) about the evolution of the H$\alpha$ emission line in the carbon
symbiotic star Draco C-1.

The intensity of the H$\alpha$ emission component of StH$\alpha$~55 appears normal
for a carbon Mira variable. Mikulasek and Graf (2005) have measured the 
integrated flux of the H$\alpha$ along the pulsation cycles of many carbon Miras.
Their results, expressed as equivalent width of the H$\alpha$ emission vs. 
the carbon Mira brightness above the minimum value, are prsented in Figure~7.
This figure shows that ($i$) the H$\alpha$ emission component can reach 
an intensity 5$\times$ stronger in field carbon Miras than seen in StH$\alpha$~55,
and ($ii$) the variability of the H$\alpha$ emission component in StH$\alpha$~55
follows the mean relation for carbon Miras.

Therefore, the Balmer emission lines observed in StH$\alpha$~55 are similar
in shape and intensity to those seen in normal carbon Miras, and therefore
do not support a binary, interactive nature of StH$\alpha$~55.

  
\vskip6mm

\References

   \refb  Arenou~F., Grenon~M., Gomez~A. 1992, A\&A 258, 104
   \refb  Barnbaum~C. 1992a, AJ 104, 1585
   \refb  Barnbaum~C. 1992b, ApJ 385, 694
   \refb  Barnbaum~C., Hinkle~K.~H. 1995, AJ 110, 805 
   \refb  Barnbaum~C., Stone~R.~P.~S., Keenan~P.~C. 1996, ApJS 105, 419
   \refb  Belczynski~K., Mikolajewska~J., Munari~U., Ivison~R.~J., Friedjung~M. 2000, A\&AS 146, 407
   \refb  Dean~C.~A. 1976, AJ 81, 364
   \refb  Deeming~T.~J. 1975, Ap\&SS 36, 137
   \refb  Downes~R.~A., Keyes~C.~D. 1988, AJ 96, 777
   \refb  Feast~M.~W., Whitelock~P., Carter~B. 1990, MNRAS 247, 227
   \refb  Feast~M.~W., Whitelock~P., Menzies~J.~W. 2006, MNRAS 369, 791
   \refb  Fiorucci~M., Munari~U. 2003, A\&A 401, 781
   \refb  Ivison~R.~J., Bode~M.~F., Meaburn~J. 1994, A\&AS 103, 201
   \refb  Jaschek~C., Jaschek~M. 1987, The Classification of Stars, Cambridge Univ. Press
   \refb  Joy~A.~H.  1954, ApJS 1, 39
   \refb  Hoffmeister~C., Richter~G., Wenzel~W. 1985, Variable Stars, Springer-Verlag
   \refb  Keenan~P.~C. 1993, PASP 105, 905
   \refb  Kerschbaum~F., Hron~J., 1992 A\&A 263, 97
   \refb  Kerschbaum~F., Groenewengen~M.~A.~T., Lazaro~C. 2006 A\&A 460, 539
   \refb  Landolt~A.~U. 1983, AJ 88, 439
   \refb  Landolt~A.~U. 1992, AJ 104, 340
   \refb  Mateo~M.~L. 1998, ARA\&A 36, 435
   \refb  Mennessier~M.-O., Boughaleb~H., Mattei~J.~A. 2000, IAUS 177, 165
   \refb  Menzies~J.~W., Feast~M.~W., Whitelock~P.~A. 2006, MNRAS 369, 783
   \refb  Mikulasek~Z., Graf~T. 2005, CoSka 35, 83
   \refb  Munari~U. 1991, A\&A 251, 103
   \refb  Munari~U., Zwitter~T. 2002, A\&A 383, 188
   \refb  Nowotny~W., Lebzelter~T., Hron~J., Hofner~S. 2005, A\&A 437, 285
   \refb  Pavlenko~Ya.~V., Marrese~P.~M., Munari~U. 2003, in Gaia Spectroscopy, Science and Technology, U.~Munari ed.,
          ASPC 298, 451
   \refb  Pojmanski~G. 1997, AcA 47, 467
   \refb  Sanford~R.~F. 1944, ApJ 99, 145
   \refb  Sanford~R.~F. 1950, ApJ 111, 270
   \refb  Soszynski~I., Dziembowski~W.~A., Udalski~A. et al.  2007, AcA 57, 201
   \refb  Stephenson~C.~B. 1986, ApJ 300, 779
   \refb  Tokunaga~A.~T., Simons~D.~A., Vacca~W.~D. 2002, PASP 114, 180
   \refb  van~Leeuwen~F. 2007, Hipparcos: The New Reduction of the Raw Data, Springer
   \refb  van~Leeuwen~F., Feast~M.~W., Whitelock~P.~A., Laney~C. 2007, MNRAS 379, 723
   \refb  van~Winckel~H., Duerbeck~H.~W., Schwarz~H.~E. 1993, A\&AS 102, 401
   \refb  Wallerstein~G.~W., Knapp~G.~R. 1998, ARA\&A 36, 369
   \refb  Whitelock~P.~A., Feast~M. Catchpole~R. 1991, MNRAS 248, 276
   \refb  Whitelock~P.~A., Feast~M.~W., Marang~F., Gronewegen~M.~A.~T. 2006, MNRAS 369, 751.
   \refb  Whitelock~P.~A., Feast~M.~W., van Leeuwen~F. 2008, MNRAS 386, 313

\end{document}